\documentclass[twocolumn,showpacs,floatfix,prl]{revtex4}

\usepackage[dvips]{epsfig}

\usepackage{float}

\begin{document}

\title{Telegraph Noise and Fractional Statistics in the Quantum Hall Effect}

\author{C.L. Kane}
\affiliation{Dept. of Physics, University of Pennsylvania, Philadelphia, PA 19104}


\begin{abstract}
We study theoretically nonequilibrium noise in the fractional quantum Hall regime
for an Aharonov Bohm ring  which has a third contact in the middle of the ring.
We show that as a consequence of their fractional statistics the tunneling of a
Laughlin quasiparticle between the inner and outer edge of the ring changes the
effective Aharonov Bohm flux experienced by quasiparticles going around the ring,
leading to a change in the conductance across the ring.  A small current in the
middle contact therefore gives rise to fluctuations in the current flowing across
the ring which resemble random telegraph noise.  We analyze this noise using the
chiral Luttinger liquid model. At low frequencies the telegraph noise varies
inversely with the quasiparticle tunneling current, and can be much larger than
the  shot noise.  We propose that combining the Aharonov Bohm effect with a noise
measurement provides a direct method for observing fractional statistics.

\end{abstract}

\maketitle

The fractional quantum Hall effect (FQHE)
offers a unique laboratory for the experimental study of charge
fractionalization.  At filling $\nu=1/m$ the FQHE state supports
quasiparticles with charge $-e/m$\cite{laughlin}. Shortly after
Laughlin's pioneering explanation of the FQHE,
Halperin\cite{halperin} pointed out that in addition to having
fractional charge, Laughlin quasiparticles (LQP's) obey fractional
statistics.  As elaborated further by Arovas, Schrieffer and
Wilczek\cite{arovas}, when two LQPs are adiabatically interchanged
in the plane, the many particle wavefunction picks up a quantum
mechanical phase $\Theta_m = \pi/m$. Equivalently, when one LQP is transported around
another a statistical phase $2 \Theta_m$ is acquired. LQPs are thus neither
bosons nor fermions, but rather more general particles, dubbed
anyons by Wilczek\cite{wilczek}.

Shot noise experiments using a quantum point contact (QPC)
setup suggested earlier\cite{kfnoise} allowed de-Piccioto et al.\cite{depiccioto}
 and Saminadayar et al.\cite{saminadayar} to
perform a direct measurement of the fractional charge
of the LQP.  To date there has been no similarly direct
observation of the fractional statistics of the LQP. Such an
observation requires a quantum interference measurement to extract
the statistical phase.  Proposed experiments have focused on the
equilibrium Aharonov Bohm (AB) effect\cite{kivelson,jain,chamon1}.
More recently, Safi et al.\cite{safi} have suggested a three
terminal Hanbury-Brown Twiss type noise experiment.  In this letter
we propose a method for directly probing the statistical phase by
combining the AB effect with a noise measurement. For our geometry
fractional statistics leads to a qualitatively new kind of noise,
which resembles random telegraph noise.

This work was inspired by a recent experiment in
which Ji et al.\cite{ji} constructed an electronic Mach-Zehnder
interferometer by fabricating a ring shaped quantum Hall sample
with QPCs along with a
lead {\it inside} the ring.  We propose a variant on
their geometry shown in Fig. 1.
QPCs 1 and 2
allow LQPs on the outer edge to circle the ring, leading to
an AB effect in the current flowing between leads 1 and 2.
QPC3 allows LQPs to
tunnel between the inner and outer edges of the ring.  Lead 3 is on the
inner edge.  In Fig. 1 we have extended lead 3 under a ``bridge".
This is useful for modeling the lead as a pair of edges
which extend to infinity.  We emphasize, however, that
this arrangement is topologically equivalent to having the contact
physically inside the ring.

\begin{figure}
 \centerline{ \epsfig{figure=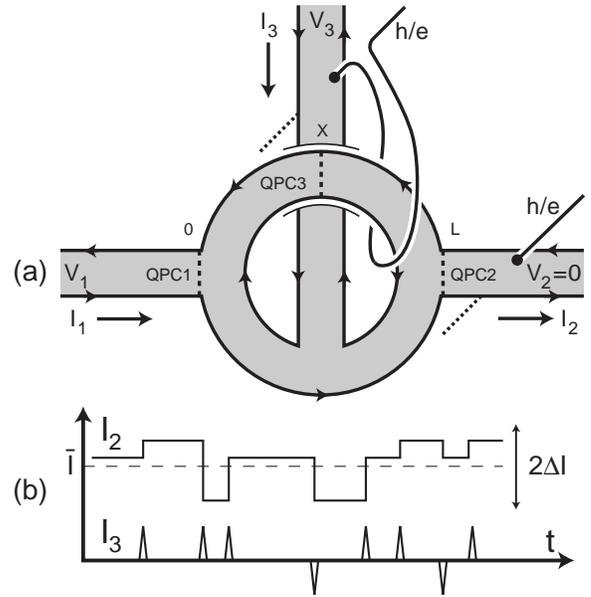,width=3.0in} }
 \caption{(a) Schematic of a three terminal AB ring
 with a third contact in the middle of the ring.  The arrows represent
 the edge state propagation and the dashed lines are QPCs.  The third
 contact has been extended under a bridge, as described in the
 text.  The dots represent LQP's, with
 flux tubes due to fractional statistics attached.
 When a LQP passes from lead 2 to lead 3 the
 effective AB flux in the ring increases by $h/e$, causing the
 current in lead 2 to switch.  (b) The resulting ``3 state" telegraph noise
 in lead 2 predicted for $\nu=1/3$ is contrasted with the shot noise in lead 3.}
 \end{figure}

The consequences of fractional charge and statistics for
equilibrium AB measurements without the middle lead are well known
\cite{kivelson,jain,chamon1,simmons,lee,goldman}.
A naive argument says that the fractional charge $-e^* = -e/m$ of the
LQP should lead to  AB
oscillations with period $h/e^* > h/e$.
This violates gauge invariance though, since a single flux quantum in the
ring can be eliminated with a gauge transformation.  The
resolution of this apparent paradox is that when the flux through
the ring is decreased by $h/e$ the ring is in an excited state with
a quasihole of charge $e^*$ on the inner edge.  In equilibrium, a
LQP from outside of the ring will eventually fill the hole.  Due to its
fractional statistics, the added LQP gives an
extra phase $2 \Theta_m$ to a  LQP circling the ring which
exactly cancels the AB phase, restoring the
$h/e$ periodicity.  Thus, the absence of multiple AB periods,
coupled with the fractional charge of the LQPs constitutes a
confirmation of the fractional statistics.  But since that absence
is guaranteed by gauge invariance, this would hardly make for a satisfying
``observation".  Chamon et al.\cite{chamon1} have argued $h/e^*$ oscillations
signifying fractional charge and statistics would be
present if the charge on the inner edge could be
appropriately adjusted with an electrostatic gate.

Gefen and Thouless\cite{gefenthouless} have argued that for a ring
with a lead in the middle an AB period $h/e^*$ could in principle
be observed provided the experiment was done on a sufficiently
fast time scale. This is a consequence of the nontrivial topology
of the structure, which has a non contractable loop enclosing a
contact.  This leads to an $m$-fold degeneracy in the FQHE
ground state\cite{thouless} that is intimately related to fractional
statistics\cite{wenniu}. Tunneling between the
degenerate states occurs when LQPs tunnel through the FQHE
fluid between the inner and outer edges of the ring.  $h/e^*$ AB
 oscillations can occur on a time scale faster than the
inverse tunneling rate.

Tunneling of LPQ's across QPC3 will lead to temporal fluctuations
in the AB current passing between leads 1 and 2. This can be seen
most transparently by  the following topological argument,
illustrated in Fig. 1a. Because of their fractional statistics,
every LQP is endowed with a ``statistical flux tube" seen by other
quasiparticles\cite{arovas}.  This represents the statistical
phase $2\Theta_m$ acquired when LQPs circle each other as an
effective AB phase, and is drawn as a vertical line passing
through the LPQ in lead 2 in Fig. 1a.  When that LPQ propagates
along the top edge to QPC3 and tunnels to the inner edge of the
ring the effective flux experienced by LPQs circling the outside
of the ring changes by $h/e$.  When the LPQ passes under the
``bridge" the statistical flux cannot pass through the FQHE state above
unless another LPQ tunnels back to the outer edge.
The flux gets ``hung up" in the ring and persists even after the
LPQ has disappeared into lead 3.  With each passing LPQ the
effective AB phase acquired by a LQP encircling the perimeter
changes by $2\Theta_m$, causing the AB current flowing between
leads 1 and 2 to switch. This leads to a pattern of
fluctuations in the current $I_2$ shown in Fig. 1b. $I_2$ switches
cyclically between $m$ values, which are characteristics of the
$m$ degenerate ground states.  These fluctuations resemble random
``$m$-state" telegraph noise.

Telegraph noise is a direct consequence of a fractional
statistical phase.  For sufficiently weak tunneling it
could in principle be observed in real time.  However, we now show
that it also has a distinct signature in the more
experimentally accessible low frequency noise.
We begin with a simple statistical argument for the noise, which
will be justified below with a detailed
calculation based on the chiral Luttinger liquid (CLL) model.  Suppose
that over a time interval $T_0$ there are $N$
tunneling events at random times $t_k$.  Tunneling
events in the forward and backward directions, denoted by $s_k
= \pm 1$, occur with probability $p^\pm$.  If the voltage
difference between the opposite edges at QPC3 is
$V_3$\cite{caveat} and the temperature is $T$, then detailed
balance dictates that $p^+ = p^- \exp(e^* V_3/T)$.  The average
tunneling current is then $I_3 = (e^* N/T_0) \tanh e^* V_3/2T$.
Here and in the following we set $\hbar=k_B=1$.

We assume that the current measured in lead 2 switches when the
LQPs tunnel, and has the simple form
\begin{equation}
I_2(t) = \bar I + \Delta I \cos\left(\phi_0 + 2 \Theta_m \sum_k s_k
\theta(t-t_k)\right).
\end{equation}
Here $\phi_0$ is an
unspecified phase, and $\theta$ is a step function.  Clearly the
average value of the current will simply be $\langle I_2 \rangle =
\bar I$.  To compute the noise we consider the correlation
function, $S(t) = \langle I_2(t_0) I_2(t_0+t) \rangle - \bar I^2$.
Averaging over the $N$ times $t_k$ and signs $s_k$, this may be
written as \begin{equation} S(t)\!=\!{\Delta I^2\over 2}\! {\rm
Re}\!\!\left[ \int_0^{T_0}\!{dt_k\over T_0}\!\!\sum_{s_k=\pm} \!\!
p^{s_k} z_m^{s_k (\theta(t_0-t_k)- \theta(t_0+t - t_k))}\right]^N,
\end{equation} where $z_m = \exp 2 i \Theta_m = \exp 2\pi i/m$ and $m\ne 1$. In the limit $N, T_0
\rightarrow \infty$ with $I_3$ fixed this becomes
\begin{equation}
S(t) = {\Delta I^2\over 2}{\rm Re}\left[e^{-{I_3 |t|\over e^*}
\left( \coth { e^* V_3 \over 2T } (1 - \cos {2\Theta_m} ) - i
\sin {2\Theta_m} \right)}\right].
\end{equation}
For low
frequency the noise will be\cite{factorof2}
\begin{equation}
S(\omega\rightarrow 0) = {e^* \Delta I^2 \over {2 I_3}} { \coth
(e^* V_3/2T) \over{ 1 + \sin^2\Theta_m/ \sinh^2(e^* V_3/2T)}}.
\end{equation}

For $e^* V_3 \gg T$, all tunneling events will be in the
forward direction, and (4) reduces to $S = e^* \Delta I^2/(2 I_3)$.
For $V_3=0$ there will be switching due to thermal fluctuations
in QPC3, and $S = {e^*}^2\Delta I^2/(4
G_3 T \sin^2\Theta_m)$, where $G_3 = I_3/V_3$ is the conductance
of QPC3.   The frequency dependence of the telegraph noise
can easily be determined from (3).  For frequency  $\omega < \omega_c = (
I_3/e^*)\coth(e^* V_3/2T)$ it will be frequency
independent and given by (4).  For $\omega > \omega_c$, $S(\omega) =
\Delta I^2/(2\omega^2)$.

In addition to this telegraph noise, there will also be shot noise
in the current $I_2(t)$.  The tunneling of LQPs through
QPC3 will generate noise of order $e^* I_3$, while backscattering
at QPC1 and QPC2 will generate noise of order $e \bar I$.  Unlike
shot noise, however, the telegraph noise varies inversely with the
tunneling current $I_3$.  Thus when $I_3 \ll \bar I, \Delta I$,
telegraph noise will be the dominant contribution to the low
frequency noise.

Eq. 4, which we established using a heuristic argument, is
our central result.  We now derive (4) within a CLL
model\cite{wen}.
In addition to providing a concrete theoretical
foundation for our assertions, this model
calculation gives predictions for the voltage ($V$) and temperature ($T$)
dependence of $\bar I$, $\Delta I$ and $I_3$.
The CLL model is a low energy theory, valid in the limit that
there is a single edge mode.  While real edges may not be in this
single channel limit, we suspect that (4) may be of more general validity,
since it reflects the topological properties of the bulk LPQ.  The detailed
$V$ and $T$ dependence however will be sensitive to the single
channel assumption.

We focus on the limit in which the LQP backscattering
at QPC3 is weak.  This is essential for observing telegraph
noise, since we require those tunneling events to be
uncorrellated.  We also consider the LQP
backscattering at QPC1 and QPC2 to be weak.  While this limit is
not necessary for experiment, it allows us to perform a perturbative
analysis of the backscattering at QPC1 and QPC2.  We compute
the currents $I_2$ and $I_3$ and the corresponding noise as
functions of voltages $V_1$ and $V_3$, with $V_2=0$.
$I_2$  will be $(e^2/mh)V_1$ with a small correction due to LPQ backscattering
at QPC1 and QPC2.  The two backscattering processes
will interfere.  Without LQP tunneling at QPC3 this leads to
AB oscillations with period $h/e^*$ in the reflected current.
We will show that LPQ tunneling at QPC3 eliminates the equilibrium AB
oscillations and leads to the telegraph noise in Eq. 4.

In the CLL model\cite{wen}, the low energy edge excitations
of the structure in Fig. 1 are described by the Hamiltonian
${\cal H} = {\cal H}_1^0 + {\cal H}_2^0 + {\cal H}_3^0 +
U_1 + U_2 + U_3$.  ${\cal H}_i^0$ describes the edge incident from
lead $i$,
\begin{equation}
{\cal H}_i^0 = {m v_F\over {4\pi}} \int dx_i \left[
\partial_x\phi(x_i)\right]^2.
\end{equation}
$v_F$ is the edge state velocity, and the coordinates $x_i$ are defined so that at QPC1 $x_i=0$, at
QPC2 $x_i = L$ and at QPC3 $x_i = X$.  The fields $\phi_i(x_i)$
satisfy $[\phi_i(x_i),\phi_j(x_j')] =
i (\pi/m) \nu_i \delta_{ij} {\rm sign}(x_i-x_j')$, where  $\nu_1 = -\nu_2 = \nu_3=1$
specifies the propagation direction.

Tunneling of LQPs at QPC$\alpha$
is described by
\begin{equation}
U_\alpha = v_\alpha O^+_\alpha e^{-i e^* \tilde V_\alpha t} + v^*_\alpha
O^-_\alpha e^{i e^*\tilde V_\alpha t},
\end{equation}
where $v_\alpha$ are the complex LQP backscattering
matrix elements at QPC$\alpha$.   The relative phase of $v_1$ and $v_2$
will depend on the
magnetic flux in the ring, advancing by $2\pi/3$ with every added
flux quantum.
The exponential factors reflect the voltage difference between the
edge states incident on the junction, $\tilde V_1 = \tilde V_2 =
V_1$, $\tilde V_3 = V_3$.
The LQP backscattering operator for QPC1 is
\begin{equation}
O^\pm_1 = {\kappa^\pm_1\over{(2\pi\eta)^{1/m}}} e^{\pm
i(\phi_1(0) - \phi_2(0))},
\end{equation}
with similar
expressions for $O^\pm_2$ and $O^\pm_3$.
$\eta$ is an ultraviolet cutoff, and  $\kappa^\pm_\alpha$ are Klein factors,
which are necessary to ensure the correct commutation relations
between the operators $O_\alpha^\pm$\cite{guyon}.  $\kappa^\pm_\alpha$ will play a crucial
role in what follows.  Their properties may be deduced from the
following physical argument.  When a LQP backscatters at
QPC1, a vortex moves the small
distance across the contact.  While this will affect the overall
phase of the wave function at another part of the sample, it can
have no immediate effect on the phase {\it difference} across
another point contact.  Different tunneling
operators must therefore {\it commute} with one another.  Using the commutation
relations obeyed by $\phi_i(x_i)$ it is then straightforward to
establish that the Klein factors must satisfy
$
\kappa_\alpha^r \kappa_\beta^s = \kappa_\beta^s \kappa_\alpha^r
e^{i r s \xi_{\alpha\beta} \pi/m}
$
with $r,s=\pm$.  Here $\xi_{\alpha\beta} = -\xi_{\beta\alpha}$ with
$\xi_{12} = 0$ and $\xi_{13} = -\xi_{23}=1$.

We begin by evaluating the average current $I_2 =  G_0 V_1 -
I^b$.  Here $G_0 = e^2/mh$ and $I^b$ is the current backscattered by QPC1 and QPC2, which
may be written as
\begin{equation}
I_b = \sum_{\alpha=1}^2 \Big\langle T_C \left[ \hat I_\alpha(t_0) e^{-i
\int_C d\tau \sum_{\beta=1}^3 U_\beta(\tau)} \right]
\Big\rangle_0.
\end{equation}
$\langle...\rangle_0$ denotes a thermal expectation value for $v_\alpha =
0$.  $\hat I_\alpha(t) = -i e^* (v_\alpha O^+_\alpha e^{-i e^* \tilde V
t} - c.c)$ are current operators.  $C$ is the
Keldysh contour, which runs from $-\infty$ to $\infty$ and back.
$T_C$ specifies time ordering on the Keldysh contour.

Consider first the case $v_3=0$\cite{chamon1}.  Expanding (8) to second order in
$v_1$ and $v_2$ and evaluating the time integrals gives
$I_2 = \bar I  + \Delta I \cos \phi_0$, where
\begin{equation}
\bar I = G_0 V_1 - e^* (|v_1|^2 + |v_2|^2)T^{2/m-1} F_1(e^* V_1/2\pi T),
\end{equation}
\begin{equation}
\Delta I = 2e^* |v_1||v_2| T^{2/m-1} F_2(e^* V_1/2\pi T,2\pi T L/v_F),
\end{equation}
and $\phi_0$ is the phase of $v_1^* v_2$.
The first term
describes backscattering from a single QPC, with
$F_1(\tilde v) = |\Gamma(1/m+i\tilde v)|^2 \sinh\pi\tilde v /(\pi
\Gamma(2/m))$.
The second term describes the interference between the two QPCs
with
\begin{equation}
F_2(\tilde v,\tilde L) = {\Gamma(1/m-i \tilde
v)(e^{2\pi\tilde v}-1)\over
{ \pi\Gamma(1/m) (2\sinh\tilde L)^{1/m}}}
Q^{i\tilde v}_{1/m-1}(\coth\tilde L).
\end{equation}
Here $Q^p_q(x)$ is the associated Legendre function of the second
kind.  $F_2$ describes  how the interference $\Delta I$ is suppressed when
$V$ or $T$ is larger than $v_F/L$.  For fixed $T \lesssim v_F/L$,
$\Delta I$ shows oscillations as a function of $V$ with a period
$2\pi v_F/e^* L$.  $\Delta I$ depends on magnetic field, and predicts
AB oscillations with period $h/e^*$.  To remedy that error
we must include $v_3$.  Though $v_3$ is small, it will
be necessary to expand (8) to all orders in $v_3$.

The perturbative expansion of (8) in powers of $v_3$ is similar to the partition
function for a collection of charged ``particles", which
correspond to tunneling events at different
times\cite{chamon}.  Forward and backward tunneling events have opposite charge.
At finite temperature, positive and negative charges will
be confined in pairs, separated by a time of order $1/T$.
Physically, these pairs correspond to real tunneling events when
they occur on opposite Keldysh paths and virtual tunneling events
when they are on the same path. When $v_3 T^{1/m-1}\ll 1$, those pairs will
be dilute and never overlap.  In that limit it is
straightforward to expand $I_b$ to all orders in $v_3$.

The $v_3^{2N}$ term in the expansion involves a set of $N$ pairs of charges.
The positive (negative) charges are at $t_k^\pm = t_k \pm \delta
t_k/2$, for $k = 1...N$.  There are two opposite
charges coming from $v_1$ and $v_2$ at
$t_0^\pm = t_0 \pm \delta t_0/2$.
To keep track of the forward
and backward paths in this expansion, we introduce an index $\sigma_k^\pm =
\pm 1$ such that $\tau_k^\pm \equiv ( t_k^\pm;\sigma_k^\pm)$\cite{kfnoise2}.
The 2Nth term in the expansion will involve an expectation value of
the form
\begin{equation}
\Pi = \Big\langle T_C[
O^+_\alpha(\tau_0^+) O^-_\beta(\tau_0^-)\prod_{k=1}^N O^+_3(\tau_k^+)
O^-_3(\tau_k^-)]\Big\rangle.
\end{equation}
This can be evaluated
as the product of two terms.  The first comes from the expectation value of the
Bose fields $\phi_i$.  When the pairs are well separated, so that $T |t_k-t_l| \gg 1$
we find that this can be written
as a product over all pairs, $\Pi_1 = G^{\sigma_0^+\sigma_0^-}_{\alpha\beta}(\delta t_0)
\prod_{k=1}^N G^{\sigma_k^+\sigma_k^-}_{33}(\delta t_k)$, where
the $G$'s are correlation functions within a single pair.
The second term
comes from the time ordered product of the Klein factors, and is given by
$
\Pi_2 = \prod_{k=1}^N z_m^{\theta(t_0-t_k) \zeta_{\alpha\beta} (\sigma_k^+ -
\sigma_k^-)/2}
$,
where $\zeta_{\alpha\beta} = (\xi_{\alpha 3}-\xi_{\beta 3})/2$.  Since $\Pi$ factorizes into a
product over $k$ of $N$ identical terms, the expansion in powers
of $v_3$ can be resummed.  Suppose we turn on $v_3$ at time $t=0$.
We then find $I_b =
\sum_{\alpha\beta} I_{\alpha\beta}$ with
\begin{equation}
I_{\alpha\beta} = I^0_{\alpha\beta} e^{- {I_3 t_0\over e^*}\left(
\coth{e^* V_3\over{2T}}
(1-\cos{2\pi\zeta_{\alpha\beta}\over m}) + i \sin{2\pi\zeta_{\alpha\beta}\over
m}\right)}.
\end{equation}
$I^0_{\alpha\beta}$ is the $v_\alpha  v_\beta^*$ term in the
expansion of (8) for $v_3=0$, and $I_3 = e^* v_3^2 T^{2/m-1}
F_1(e^* V_3/2\pi T)$.  This shows that the interference terms
average away after a time $\tanh(e^* V_3/2T)/I_3$ due to LQP
tunneling, leaving only the diagonal terms, $I_2 = \bar I$.

The noise in lead 2 is calculated by computing
$S(t) = \langle \{I_2(t_0+t), I_2(t_0)\}\rangle/2-\bar I^2$.
There will be several contributions.  At order $v_{1,2}^2$ there will be the
``single barrier" shot noise\cite{kfnoise} produced by
QPCs 1 and 2 given by $e^* I_b$ for $e^* V_3 \gg T$.
Shot noise produced by QPC3 will also be present at order $v_1^2 v_3^2$.
The telegraph noise comes in at
order $v_1^2 v_2^2$.  The $v_3$ expansion is similar to
that above, and leads precisely to (4), which is of order
$v_1^2 v_2^2/v_3^2$.

Optimal experimental
conditions for observing telegraph noise include,  (1)
$I_3 \ll G_0 V_3$, so that LQPs are dilute.
(2) $I_3 \ll \Delta I$, so that telegraph noise dominates
shot noise.  It is desirable to set QPC1 and QPC2 so that
they backscatter moderately to maximize the interference signal.
It is also necessary that (3) $V_1,T \lesssim v_F/L$, or else
 the interference will be washed out.  Finally, the
frequency of the measured noise should be low enough so that
(4) $\omega \lesssim (I_3/e^*) \coth e^* V_3/2 T$.

We now comment briefly on the generalization of the above results to
hierarchical states.  For $\nu=1/m$ the quasiparticle charge,
statistics and Hall conductivity all have the same numerical value, and can be
argued to reflect the same quantity.  By contrast, for
hierarchical states these quantities differ.  For example $\nu=2/7$ has
quasiparticles with charge $-e/7$ and statistics $\Theta = 3\pi/7$.  Both the
heuristic arguments and the CLL theory can be generalized to account for
hierarchical states, and Eq. (4) is recovered with the appropriate statistics
angle $\Theta$.

It is a pleasure to thank Moty Heiblum for helpful discussions.


\begin{thebibliography}{10}
\bibitem{caveat}  When backscattering at QPC2 is not weak,
this will be the difference between $V_3$ and an appropriate
average of $V_1$ and $V_2$.
\bibitem{factorof2}  This differs by a factor of 2 from
the conventional definition, which for classical shot noise gives $S =
2eI$.  With our definition $S = eI$.
\end{thebibliography}
\end{document}